\documentclass[prl,twocolumn,showpacs,preprintnumbers,amsmath,amssymb]{revtex4}

\newcommand{\hi}{{\cal{H}}}
\newcommand{\ma}{{\cal{M}}}
\newcommand{\dom}{{\cal{D}}}

\usepackage{dcolumn}
\usepackage{bm}

\begin{document}

\title{A Covariant Information-Density Cutoff in Curved Space-Time}

\author{Achim Kempf}
\affiliation{
Department of Applied Mathematics, University of Waterloo\\
Waterloo, Ontario N2L 3G1, Canada}


\begin{abstract}
In information theory, the link between continuous information and
discrete information is established through well-known sampling theorems.
Sampling theory explains, for example, how frequency-filtered music
signals are reconstructible perfectly from discrete samples. In this
Letter, sampling theory is generalized to pseudo-Riemannian manifolds.
This provides a new set of mathematical tools for the study of space-time
at the Planck scale: theories formulated on a differentiable space-time
manifold can be completely equivalent to lattice theories. There is a
close connection to generalized uncertainty relations which have appeared
in string theory and other studies of quantum gravity.
\end{abstract}


\maketitle It is generally assumed that the notion of distance loses
operational meaning at the Planck scale, $l_{P} \approx 10^{-35}m$
(assuming 3+1 dimensions), due to the combined effects of general
relativity and quantum theory. Namely, if one tried to resolve a spatial
region with an uncertainty of less than a Planck length, then the
corresponding momentum uncertainty should randomly curve and thereby
significantly disturb the very region in space that was meant to be
resolved. It is expected, therefore, that the existence of a smallest
possible length, area or volume, at the Planck scale or above, plays a
central role in the yet-to-be-found theory of quantum gravity.

In the literature, no consensus has been reached as to whether this
implies that space-time is discrete. On the one hand, quantization
literally means discretization, and space-time discreteness is indeed
naturally accommodated within the functional analytic framework of quantum
theory, see, e.g., \cite{foam}. Also, most interacting quantum field
theories are mathematically well-defined only on lattices. On the other
hand, within the mathematical framework of general relativity, space-time
is naturally described as a differentiable manifold and deep principles
such as local Lorentz invariance would appear to be violated if space-time
were discrete.

There is the possibility that the cardinality of space-time is between
discrete and continuous, but it is strongly restricted by results of
G{\"o}del and Cohen. In \cite{goedel-cohen}, they proved that both are
consistent with conventional (ZF) set theory: to adopt an axiom claiming
the existence of sets with intermediate cardinality or to adopt an axiom
claiming their non-existence. Therefore, it is not possible to explicitly
construct any set of cardinality between discrete and continuous infinity
from the axioms of conventional set theory. If space-time is describable
as a set and if this set is of intermediate cardinality, then its
description cannot be constructive and requires mathematics beyond
conventional set theory.

In this Letter, we consider a simpler possibility. In a concrete sense,
space-time could be simultaneously discrete \it and \rm continuous.
Namely, in the simplest case, physical fields could be differentiable
functions which possess merely a finite density of degrees of freedom. If
such a field's amplitude is sampled at discrete points of the space-time
manifold then the field's amplitude at all points in the manifold are
reconstructible from those samples - if the sample points are spaced
densely enough. The minimum average sample density which allows the
reconstruction of fields could be, for example, on the order of the Planck
scale. All physical entities such as fields, Lagrangians and actions can
then be written either as living on a differentiable manifold, thereby
displaying external symmetries or, equivalently, as living on any one of
the sampling lattices of sufficiently small average spacing, thereby
displaying ultraviolet finiteness. Such theories need not break any
symmetries of the manifold because among all sufficiently tightly spaced
lattices no particular lattice is preferred.

In the information theory community, the mathematics of classes of
functions which can be reconstructed from discrete samples is well-known,
namely as sampling theory. Shannon, in his seminal work \cite{shannon},
introduced sampling theory as the link between continuous information and
discrete information. Our aim here is to extend this link between discrete
and continuous information to curved space-times.

The use of sampling theory in quantum mechanics was first suggested in
\cite{ak-prl}, where a close connection was shown to generalized
uncertainty relations that had appeared in studies of quantum gravity and
string theory, see \cite{ucrs}. In the simplest case, these uncertainty
relations are of the form $\Delta x\Delta p \ge \frac{\hbar}{2}(1+\beta
(\Delta p)^2)$ and imply the existence of a finite minimum uncertainty in
position $\Delta x_{min}=\hbar\sqrt{\beta}$, as is easily verified. In
\cite{ak-prl}, it was shown that a finite lower bound to the position
uncertainty implies that the wave functions possess the sampling property,
i.e. that they can be reconstructed everywhere from discrete samples if
those samples are taken at a spacing that is at least as small as the
minimum position uncertainty. In \cite{ak-gauge}, it was suggested that
the freedom of choice of sampling lattice may be related to gauge
symmetries.

In recent proceedings, see \cite{ak-guelph}, sampling theory for physical
theories on curved spaces was outlined. Building on \cite{ak-guelph}, this
Letter introduces sampling theory on Riemannian and pseudo-Riemannian
manifolds. We will obtain a covariant information density cutoff together
with a new sampling theoretic principle that is consistent with the
Lorentz contraction of sampling lattices.


The basic sampling theorem goes back to Cauchy in the early 19th century,
see \cite{ferreira}. Consider the set of square integrable functions $f$
whose frequency content is bounded by $\omega_{max}$, i.e., which can be
written as $f(x)=\int_{-\omega_{max}}^{\omega_{max}} \tilde{f}(\omega)
e^{i\omega x} d\omega$. These $f$ are called bandlimited functions with
bandwidth $\omega_{max}$. If the amplitudes $\{f(x_n)\}$ of such a
function are known at equidistantly-spaced discrete points $\{x_n\}$ whose
spacing is $\pi/\omega_{max}$, then the function's amplitudes $f(x)$ can
be reconstructed for all $x$. The reconstruction formula is:
\begin{equation}
f(x)~ =~ \sum_{n=-\infty}^\infty ~f(x_n) ~
\frac{\sin[(x-x_n)\omega_{max}]}{(x-x_n)\omega_{max}}\label{basic}
\end{equation}
This sampling theorem is in ubiquitous use, e.g., in digital audio and
video as well as in scientific data taking. Sampling theory, see
\cite{ferreira}, studies generalizations of the theorem for various
classes of functions, for non-equidistant sampling, for multi-variable
functions and it investigates the stability of the reconstruction in the
presence of noise.

Following \cite{ak-guelph}, we now define a general framework for sampling
on Riemannian manifolds. The key assumption in the basic sampling theorem
is a frequency cutoff, more precisely a cutoff of the spectrum of the
self-adjoint differential operator $-id/dx$. On a multi-dimensional curved
space, a covariant analog of a bandlimit is the cutoff of the spectrum of
a scalar self-adjoint differential operator. As an explicit example one
may choose the Laplace-Beltrami operator $\Delta = \vert
g\vert^{-1/2}\partial_ig^{ij}\vert g\vert^{1/2}\partial_j$ where $\vert
g\vert$ is the determinant of the metric tensor $g$.

Consider then the Hilbert space $\hi$ of square integrable, say scalar,
functions over the manifold and the dense domain $\dom\subset\hi$ on which
the considered operator, say the Laplacian, is essentially self-adjoint.
Using sloppy but convenient terminology we will speak of all points of the
spectrum as eigenvalues, $\lambda$, with corresponding ``eigenvectors"
$\vert \lambda)$, keeping in mind that the manifold will generally be
noncompact and its spectrum therefore not discrete. We use the notation
$\vert ~)$ in analogy to Dirac's bra-ket notation, but with round brackets
to distinguish from quantum states. The $\vert \phi)$ that we consider
here could be, for example, the scalar fields that are being integrated
over in a quantum field theoretical path integral.

The operator $-\Delta$ is positive and its spectrum is an invariant of the
manifold. A spatially covariant ``bandwidth" cutoff in nature then means
that physical fields are elements of $\dom_{ph}= P.\dom$, where $P$
projects onto the subspace of $\dom$ which is spanned by the eigen\-spaces
of $-\Delta$ with eigenvalues smaller than some fixed maximum value
$\lambda_{max}$, which could be, e.g., on the order of $1/l^2_P$.

For example, in quantum field theoretical actions this type of cutoff
arises if $-\Delta$ is the lowest order term in a power series in
$-\Delta$ whose radius of convergence is finite, say $1/l^2_P$. Examples
are the geometric series $l^{-2}_P\phi^*\sum_{n=1}^\infty
(-l^2_P\Delta)^n\phi$ and $l^{-2}_P\sum_{n=1}^\infty
(-l^2_P\phi^*\Delta\phi)^n$. Such series correspond to Planck-scale
modified dispersion relations, a concept that has recently attracted
considerable attention in the context of the transplanckian problems in
black hole radiation and inflationary fluctuations, see, e.g.,
\cite{unruhetc,cosm}. Interestingly, also the Dirac-Born-Infeld action may
be viewed as providing a minimum length cutoff through this mechanism,
namely when expanding the square root in its action as power series with
finite radius of convergence. See, in particular, \cite{schuller}.

If, by this or another mechanism, the yet-to-be-found theory of quantum
gravity does yield a bandwidth cutoff, how do the fields in the physical
domain $\dom_{ph}$ acquire the sampling property? For simplicity, assume
that one chart covers the $N$-dimensional manifold. The coordinates
$\hat{x}_j$, for $j=1,...,N$ act as multiplication operators that map
scalar functions to scalar functions: $\hat{x}_j: \phi(x) \rightarrow x_j
\phi(x)$. On their domain within the Hilbert space $\hi$ these operators
are essentially self-adjoint, with an ``Hilbert basis" of non-normalizable
joint eigenvectors $\{\vert x)\}$ with continuum normalization ${\bf 1
}=\int d^Nx~\vert g\vert^{1/2} ~\vert x)(x\vert$. We have $( x\vert
\phi)=\phi(x)$. Since $\Delta$, being a differential operator, cannot
commute with the position operators $\hat{x}_j$ we obtain the situation
$P\vert x)\neq \vert x)$. Thus, on the restricted domain $\dom_{ph}$, the
multiplication operators $\hat{x}_j$ are merely symmetric but not
self-adjoint (intuitively, for lack of eigenvectors). Correspondingly, the
uncertainty relations are modified, similar to the toy cases discussed in
\cite{ak-old}.

Consider now a physical field, i.e., a vector $\vert \phi)\in \dom_{ph}$.
Assume that the field's amplitudes $\phi(x_n)=( x_n\vert \phi)$ are known
at discrete points $\{x_n\}$ of the manifold. While all position
eigenvectors $\vert x)$ are needed to span $\hi$, sufficiently dense
discrete subsets $\{\vert x_n)\}$ of the set of vectors $\{P\vert x)\}$
can span $D_{ph}$. A field's coefficients $\{\phi(x_n)\}$ then fully
determine the Hilbert space vector $\vert \phi)\in D_{ph}$ and they
determine, therefore, also $(x\vert\phi)$ for all $x$. Namely, defining
$K_{n\lambda}= (x_n\vert\lambda)$, the set of sampling points $\{x_n\}$ is
sufficiently dense for reconstruction iff $K$ is invertible. To see this,
insert the resolution of the identity in terms of the eigenbasis
$\{\vert\lambda )\}$ of $-\Delta$ into $(x\vert\phi)$:
\begin{equation}
(x\vert\phi)=\sum_{\vert\lambda\vert<\lambda_{max}}
\!\!\!\!\!\!\!\!\!\!\!\!\!\!\int ~(x\vert \lambda)(\lambda \vert
\phi)~d\lambda \label{abc}
\end{equation}
We use the combined sum and integral notation since the spectrum of
$-\Delta$ may be discrete and/or continuous (the manifold $\ma$ need not
be compact) and it is understood that eigenvalues can be degenerate. With
$K$ invertible, one obtains $(\lambda\vert \phi)=\sum_n K^{-1}_{\lambda,n}
\phi(x_n)$ which, when substituted back into Eq.\ref{abc}, yields
\begin{equation}
\phi(x) = \sum_n G(x,x_n) ~\phi(x_n),\label{recon}
\end{equation}
with the reconstruction kernel:
\begin{equation}
G(x,x_n)~=~\sum_{\vert\lambda\vert<\lambda_{max}}
\!\!\!\!\!\!\!\!\!\!\!\!\!\!\int ~ (x\vert\lambda) K^{-1}_{\lambda,
n}~d\lambda\label{recon2}
\end{equation}
In conventional applications of information theory, the sample points
$\{x_n\}$ are required to be dense enough to allow stable reconstruction
in the presence of noise: functions reconstructed from small samples must
have small norm, in the sense that there exists a $C>0$ such that
\begin{equation}
(\phi\vert\phi)~\le~ C \sum_n \vert\phi(x_n)\vert^2~~~\text{for
all}~\vert\phi)\in D_{ph}. \label{frame}
\end{equation}
The minimum sample density for stable reconstruction in flat Euclidean
space equals the bandwidth volume in Fourier space, up to a constant, as
was shown by H.J. Landau in \cite{landau}. It is also the density of
degrees of freedom, defined as the dimension of the space of bandlimited
functions which possess essential support in a given volume.
Reconstruction stability is highly nontrivial already for classical
signals in flat space: a bandwidth cutoff does not prevent bandlimited
functions from oscillating arbitrarily fast in an arbitrarily large
region. These superoscillations recruit degrees of freedom from outside
the considered region at the expense of reconstruction stability. As was
shown in \cite{ak-super-prl}, superoscillations in quantum mechanical wave
functions produce effects that raise thermodynamic and measurement
theoretic issues. It should be most interesting, therefore, to explore for
physical fields in curved space-time the role of quantum fluctuations as
noise and to use our new approach to generalize Landau's theorem to
Riemannian manifolds. For fixed noise, the density of degrees of freedom
then yields the maximum Shannon information density as usual, see
\cite{shannon}.

We close the case of Riemannian manifolds with several remarks. In the
case of one dimension and equidistant samples, Eqs.\ref{abc}-\ref{recon2}
reduce to the basic sampling theorem of Eq.\ref{basic}. Another simple
case is that of compact manifolds. Their Laplacian possesses a discrete
spectrum whose cutoff renders $\dom_{ph}$ finite-dimensional. A
corresponding finite number of sampling points suffices for
reconstruction. Since sampling theory has its origins and most of its
applications in communication engineering, sampling theory on Riemannian
manifolds has been little studied so far. An exception is the $SU(2)$
group manifold, for which the spectral cutoff yields the much-discussed
fuzzy sphere, see, e.g., \cite{madore}. Very interesting results were
obtained by Pesenson, see, e.g., \cite{pesenson}, who considered, in
particular, the case of homogeneous manifolds. In \cite{pesenson},
reconstruction works differently, however, namely by approaching the
solution iteratively in a Sobolev space setting. Useful methods should
also be available from the field of spectral geometry, see, e.g.,
\cite{specs1}, which studies the close relationship between the properties
of a manifold and the spectrum of its Laplacian and, in particular, from
the field of noncommutative geometry and the techniques based on the
spectral triple, see \cite{connesmajid}.

Let us now turn to the entirely new case of sampling in pseudo-Riemannian
manifolds. We define a (N+1) dimensional covariant ``bandlimit" as a
cutoff of the spectrum of a scalar self-adjoint differential operator such
as the Dirac or the d'Alembert operator $\Box$ (as opposed to, e.g.,
regularization through a generalized $\zeta$-function). While these
operators are self-adjoint, they are not elliptic and their spectrum needs
to be cut off from above and below. This will lead us to a new
sampling-theoretic principle that accounts for the Lorentz contraction of
lattice spacings: Each temporal frequency component
$\phi(\omega,\underline{x})$ of a field possesses its own finite spatial
bandwidth and can be reconstructed from discrete spatial samples of
corresponding density. Equivalently, each spatial mode
$\phi(t,\underline{k})$ possesses a finite temporal bandwidth and can be
reconstructed from samples taken at correspondingly dense discrete
sampling times.

To see this, consider first the case of flat (N+1)-dimensional Minkowski
space-time. Fourier theory is applicable and cutting off the spectrum of,
e.g., the d'Alembertian amounts to requiring $\vert p_0^2-\vec{p}^2\vert <
l_{P}^{-2}$, as is implementable in the action, e.g., through a power
series in $p_0^2-\vec{p}^2$ with radius of convergence $l_{P}^{-2}$. Thus,
for each fixed $p_0$ there is a finite bandwidth volume, $B(p_0)$, in the
$N$-dimensional space of $\vec{p}$ values. The bandwidth volume is of ball
shape if $\vert p_0\vert\le l^{-1}_{P}$ and is of spherical shell shape if
$\vert p_0\vert> l^{-1}_{P}$. Each frequency component $p_0$ of a field
possesses its own finite spatial bandwidth and can be reconstructed from
discrete samples:
\begin{equation}
\phi(p_0,\vec{x}) =
\sum_{\vec{n}}\phi(p_0,\vec{x}_{\vec{n}}(p_0))G(p_0,\vec{x},\vec{x}_{\vec{n}}(p_0))
\end{equation}
For example, a kernel, $G$, for stable reconstruction from equidistant
samples follows from the sampling theorem of Eq.\ref{basic} by viewing
each of the bandwidth volumes as contained in a rectangular bandwidth box.
By Landau's theorem, \cite{landau}, the precise minimum sample density for
stable reconstruction of a temporal frequency mode, $p_0$, is given by the
bandwidth volume $B(p_0)$ in momentum space. For sub-Planckian frequency
modes, $\vert p_0\vert\le l^{-1}_P$, the maximal bandwidth volume $B(p_0)$
is the volume of the $D$-dimensional sphere with radius $\sqrt{2}
l^{-1}_P$. Thus, all these modes can be reconstructed stably from samples
spaced at around the Planckian density. For trans-Planckian frequency
modes, $\vert p_0 \vert \gg l^{-1}_{P}$, the scaling of the bandwidth
volume depends on $D$, namely $B(p_0) = O(\vert p_0\vert^{D-2})$ for
$\vert p_0\vert \rightarrow \infty$. Thus, by Landau's theorem, the
minimum sample density for stable reconstruction as $\vert p_0\vert
\rightarrow \infty$ is decreasing for $D=1$, while constant for $D=2$ and
increasing for $D\ge 3$. Therefore, in one and two spatial dimensions,
Planck scale sample density suffices to stably reconstruct all temporal
frequency modes, i.e. the entire field. In the curved space setting this
might apply, e.g., to strings/string bits and to the holographic principle
for horizons respectively, see, e.g., \cite{bekensteinetc}. For $D\ge 3$
no single sample density suffices to stably reconstruct simultaneously all
temporal modes. Similar to the discussion above, it is clear that each
spatial mode $\vec{p}$ possesses a finite temporal bandwidth and that,
independently of $D$, all spatial modes can be stably reconstructed from
times series of Planck scale spacing. Generally, if the reconstruction is
not required to be stable, significantly sparser sample densities may
suffice.

The analysis of Minkowski space generalizes straightforwardly to static
space-times. These possess coordinates in which the d'Alembertian consists
of a simple temporal part and an elliptic self-adjoint spatial part. In
this case, each temporal frequency mode possesses the equivalent of a
spatial bandwidth, and vice versa. It should be most interesting to carry
through a corresponding analysis of the sampling theory in generic
space-times with singularities and with horizons, where space-like and
time-like coordinates can switch their roles. Recall that while the
reconstruction formulas of fields depend on the choice of coordinates, the
physics of the bandwidth cutoff is, of course, invariant. For those
pseudo-Riemannian manifolds which allow Wick rotation the situation
reduces to the simpler case of Riemannian manifolds and Wick rotation of
the reconstruction formulas.

Of interest to phenomenology is that, in typical inflationary scenarios,
modes which are today of cosmological size grew from the Planck scale
merely about five orders of magnitude before their dynamics froze upon
crossing the Hubble horizon. Therefore, Planck scale physics could have a
small but potentially measurable effect on the inflationary predictions
for the CMB amplitude- and polarization spectra, see \cite{cosm0}. A small
effect is predicted, in particular, if the above-mentioned generalized
uncertainty relations hold, see \cite{cosm}, which would give fields the
sampling property. So far, these models introduce Planck scale physics in
ways that are tied to the preferred foliation of space-time into the
essentially flat space-like hypersurfaces defined by the cosmic microwave
background rest frames, and these models are, therefore, breaking general
covariance. The new approach here can be used to predict possible effects
of a generally covariant sampling theoretic cutoff on CMB predictions.
While free field theory suffices for this purpose, note that field
theoretic interaction terms, i.e., higher than second powers of fields
(second powers occur as scalar products in the Hilbert space of fields)
would have to be nontrivial in order to yield a result within the cut-off
Hilbert space.

Interestingly, regularization of field theories in curved space is known
to induce in the action a series in the curvature tensor, see
\cite{sakharov}. We now see that if this series possesses a finite radius
of convergence it induces a bandwidth cutoff and sampling theorem for the
metric itself, i.e., a cutoff on the curvature of space-time.

\bf Acknowledgement: \rm ~~The author thanks F. Schuller, H. Burton, I.
Pesenson and L. Kauffman for useful comments. This work has been partially
supported by the National Science and Engineering Research Council of
Canada.

\vfill
\end{document}